\documentclass[a4paper,12pt, epsfig]{article}
\usepackage{amscd,verbatim}
\usepackage[all]{xy}
\usepackage{graphicx}

\usepackage{amsmath}
\usepackage[psamsfonts]{amssymb}
\usepackage{amsthm}
\usepackage{euscript}

\usepackage{latexsym}

\renewcommand{\(}{\begin{equation}}
\renewcommand{\)}{end{equation} \vspace{-.05in}\linebreak}
\newcommand{\bea}{\begin{eqnarray}}
\newcommand{\eea}{\end{eqnarray}}

\newcounter{saveeqn}
\newcounter{savealpheqn}

\newcommand{\alpheqn}{\setcounter{saveeqn}{\value{equation}}%
      \stepcounter{saveeqn}\setcounter{equation}{0}%
      \renewcommand{\theequation}{\mbox{\arabic{section}.\arabic{saveeqn}
\alph{equation}}}
      \renewcommand{\)}{\end{equation}}}
\def\part#1{\frac{\partial}{\partial{#1}}}%
\def\group#1{\refstepcounter{equation}\setcounter{saveeqn}{\value{equati
on}}%
      \label{#1}\setcounter{equation}{0}%
\renewcommand{\theequation}{\mbox{\arabic{section}.\arabic{saveeqn}
\alph{equation}}}
      \renewcommand{\)}{\end{equation}}}
\newcommand{\reseteqn}{\setcounter{equation}{\value{saveeqn}}%
      \renewcommand{\theequation}{\arabic{section}.\arabic{equation}}%
      \renewcommand{\)}{\end{equation}}}

\newcommand{\aalpheqn}{\setcounter{saveeqn}{\value{equation}}%
      \stepcounter{saveeqn}\setcounter{equation}{0}%
      \renewcommand{\theequation}{\mbox{
            \Alph{subsection}.\arabic{saveeqn}\alph{equation}}}
       \renewcommand{\)}{\end{equation}}}
\newcommand{\areseteqn}{\setcounter{equation}{\value{saveeqn}}%
      \renewcommand{\theequation}{\Alph{subsection}.\arabic{equation}}%
      \renewcommand{\)}{\end{equation}}}

\renewcommand{\thefootnote}{\alph{footnote}}
\renewcommand{\(}{\begin{equation}}
\renewcommand{\)}{\end{equation}}
\newcommand{\ba}{\begin{eqnarray}}
\newcommand{\ea}{\end{eqnarray}}

\newcommand{\bp}{\mathop{\vtop{\ialign{##\crcr

$\hfil\displaystyle{}\hfil$\crcr\noalign{\kern-13pt\nointerlineskip}
       \BIG{(}\hskip0pt\crcr\noalign{\kern3pt}}}}}
\newcommand{\cbp}{\mathop{\vtop{\ialign{##\crcr

$\hfil\displaystyle{}\hfil$\crcr\noalign{\kern-13pt\nointerlineskip}
       \BIG{)}\hskip0pt\crcr\noalign{\kern3pt}}}}}
\newcommand{\pa}{\mathop{\vtop{\ialign{##\crcr

$\hfil\displaystyle{\oplus}\hfil$\crcr\noalign{\kern+1pt\nointerlineskip
}
       \hspace{.08in}$^{\alpha=0}$\hskip6pt\crcr\noalign{\kern3pt}}}}}

\newcommand{\p}{^\prime}

\newcommand{\rank}{{\textup{\scriptsize{rank}}}}

\newcommand{\cE}{\ensuremath{\mathcal E}}
\newcommand{\cF}{\ensuremath{\mathcal F}}
\newcommand{\cG}{\ensuremath{\mathcal G}}
\newcommand{\cH}{\ensuremath{\mathcal H}}

\newcommand{\cK}{\ensuremath{\mathcal K}}
\newcommand{\cL}{\ensuremath{\mathcal L}}
\newcommand{\cM}{\ensuremath{\mathcal M}}

\newcommand{\h}{\mathfrak{h}}
\newcommand{\e}{\mathfrak{e}}

\newcommand{\C}{\ensuremath{\mathbb C}}

\newcommand{\Q}{\ensuremath{\mathbb Q}}

\newcommand{\Z}{\ensuremath{\mathbb Z}}

\def\dwn{\downarrow}

\newcommand{\beq}{\begin{equation}}
\newcommand{\eeq}{\end{equation}}

\newcommand{\iso}{\cong}

\numberwithin{equation}{section}

\def\hsp#1{\hspace{#1in}}

\catcode`\@=11
\def\vereq#1#2{\lower3pt\vbox{\baselineskip1.5pt \lineskip1.5pt
\ialign{$\m@th#1\hfill##\hfil$\crcr#2\crcr\sim\crcr}}}
\catcode`\@=12

\makeatletter
\newcommand\figcaption{\def\@captype{figure}\caption}
\newcommand\tabcaption{\def\@captype{table}\caption}
\makeatother
\renewcommand{\(}{\begin{equation}}
\renewcommand{\)}{\end{equation}}

\newcommand{\CC}{{\mathbb C}}

\theoremstyle{plain}

\theoremstyle{definition}

\begin{document}

\begin{titlepage}
\begin{flushright}
hep-th/0608190 \\
\end{flushright}

\vspace{2em}
\def\thefootnote{\fnsymbol{footnote}}



\begin{center}
{\large\bf  $E\sb8$ Gauge Theory and Gerbes in String Theory}
\end{center}
\vspace{1em}
\begin{center}
\Large Hisham Sati \footnote{E-mail:
hisham.sati@yale.edu}
\end{center}

\begin{center}
\vspace{1em}
{\em { Department of Mathematics\\
Yale University\\
New Haven, CT 06520\\
USA\\
\hsp{.3}\\

Department of Pure Mathematics\\
       University of Adelaide\\
       Adelaide, SA 5005\\ 
Australia\\

\hsp{.3}\\
The Erwin Schr\"odinger International\\
Institute for Mathematical Physics,\\
Boltzmanngasse 9, A-1090 Wien\\
Austria}}\\

\end{center}

\vspace{0.5cm}

\begin{abstract}
The reduction of the $E_8$ gauge theory to ten dimensions leads to 
a loop group, which in relation to twisted K-theory has a Dixmier-Douady 
class identified with the Neveu-Schwarz H-field. We give an interpretation 
of the degree two part of the eta-form by comparing the adiabatic limit 
of the eta invariant with the one loop term in type IIA. More generally, 
starting with a $G$-bundle, the comparison for manifolds with String 
structure identifies $G$ with $E_8$ and the representation as the 
adjoint, due to an interesting appearance of the dual Coxeter number. This 
makes possible a description in terms of a 
generalized WZW model at the critical level. We also discuss the relation 
to the index gerbe, the possibility of obtaining such bundles from loop 
space, and the symmetry breaking to finite-dimensional bundles. 
We discuss the implications of this and we give several proposals. 
\end{abstract}

\vfill

\end{titlepage}
\setcounter{footnote}{0}
\renewcommand{\thefootnote}{\arabic{footnote}}

\pagebreak

\renewcommand{\thepage}{\arabic{page}}

\tableofcontents

\section{Introduction}
The form-fields in M-theory and string theory play an important role in
the characterization of the global structure of the theory. The study of 
their quantization conditions and partition functions has led to 
a wealth of topological and global analytic information about the 
objects and the fields of string theory. In particular, Diaconescu,
Moore and Witten (DMW) \cite{DMW} initiated the comparison of the 
partition function in M-theory, described by index theory of an $E_8$ 
bundle and a Rarita-Schwinger bundle, with the partition function in
type IIA string theory described by K-theory.

\vspace{3mm}
We focus on the $E_8$ principal bundle \cite{Flux} on the 
eleven-dimensional spacetime of M-theory, 
\(
\begin{matrix}
E_8&\to & P\cr
&&\dwn\cr
S^1& \to & Y^{11}\cr
&&\dwn {\small{\pi}}\cr
     & & X^{10}\cr
\end{matrix}
\)
where $Y^{11}$ in turn is a principal $S^1$ bundle over the
$10$-dimensional manifold $X^{10}$. Corresponding to $P$ is an
associated vector bundle $V$, both characterized by a degree four 
integral class $a$. 

\vspace{3mm}
In \cite{DMW}, the NSNS $B$-field was switched off and so it was assumed
that the M-theory $C$-field $C_3$ is a pullback from $X^{10}$.
The implication is that the topological invariant, the phase
$\Omega_M (C_3)$, depends only on $a$ and not on $C_3$, so that one 
writes $\Omega_M (a)$. In \cite{MS1} the generalization of this to 
include the NSNS $H$-field was considered, generalizing also the 
case $[H]=0$ \cite{MS},  and corresponding to the situation
when the bundles in M-theory are not lifted from the Type IIA base.
As in \cite{MS1} our main focus will be the $E_8$ gauge theory because
the Rarita-Schwinger bundle involves only natural bundles and such
bundles are automatically lifted from the base of the $S^1$-bundle.

\vspace{3mm}
In gauge theory, it has been known that periodic instantons of a gauge 
theory with structure group $G$ on a space $Y^4=S^1 \times X^3$ give 
rise to monopoles on $X^3$ with structure group the Kac-Moody group 
of $G$ \cite{GM}. The situation in M-theory is analogous and so one 
expects that starting from an $E_8$ gauge theory on $Y^{11}$ one gets 
an ${LE}_8$ bundle on $X^{10}$ \cite{AE}. Indeed the 
computations at the classical level further confirm this 
\cite{Jarah,BV}. 
 
\vspace{3mm}
In \cite{MS1} an expression was found for the phase of M-theory by using 
the adiabatic limit of the eta invariant.
\footnote{We note (with V. Mathai) that the analysis of the phase is 
where the nontrivial circle bundle matters in \cite{MS1}. For the other 
parts of that paper we might as well have assumed a trivial circle 
bundle.}
 The resulting expression was 
an integral over the ten-dimensional base of the circle bundle, thus 
relating the M-theory data on the nontrivial circle bundle to the data 
of type IIA on $X^{10}$. However, that expression involved the eta 
forms, the higher degree analogs of the more familiar eta invariant, and 
the expression was not evaluated. There the desire was expressed to find 
an interpretation of the components of the eta form. It is the purpose of 
this note to propose such an interpretation for the first nontrivial 
eta forms, ${\widehat{\eta}}^{(2)}$ of degree two. We do this by 
comparing the expression of the adiabatic limit with the one loop term 
in type IIA. 

\vspace{3mm}
The two faces of the cohomology class in $H^3(X^{10};\Z)$, the Dixmier-Douady 
invariant, are utilized, the first being the obstruction to 
replacing the bundle 
$LE_8$ associated with a projective representation by a vector bundle
which is
the central 
extension ${\widehat{LE_8}}$, and the second is that the Dixmier-Douady 
class describes a (stable) equivalence class of a gerbe 
\footnote{We comment on the use of gerbes as geometric objects in our 
discussion. To describe higher degree fields one can also use 
differential
characters. While they are closely
related, what we have seen in the literature is that gerbes seem to be
more adapted to analytical descriptions such as index theory -- which we
use in this note-- whereas differential characters seem to be more 
useful in
the description of the gauge fields in quantum (higher) gauge theories.}
over $X^{10}$.  
This is used in order to relate the $E_8$ gauge theory to twisted 
K-theory. The former results in the loop group bundle upon reduction to 
ten dimensions and the latter can be interpreted as the K-theory of 
(bundle) gerbes \cite{BCMMS}. For applications of bundle gerbes and 
DD-classes in families problems in QFT see \cite{CMM2, CMM}.

\vspace{3mm}
Starting with the principal $E_8$ bundle, we compare the adiabatic limit 
with the one loop term in type IIA string theory in ten dimensions.
The observation is that for string manifolds, i.e. for those with 
$\lambda=p_1/2$ zero, the two-form component of the eta-form is 
identified with the NSNS gerbe. What is interesting is the integer 
multiplying the generator. This leads to an interesting 
appearance of the dual Coxeter number of $E_8$ in front of the degree 
three generator. In fact the analysis can be made general and can be 
seen, in some sense, as a {\it discovery} of $E_8$. Starting with a $G$-bundle and 
performing the dimensional reduction one gets a bundle of $LG$, the 
DD class of which is the obstruction to lifting to the bundle with 
structure group the central extension $\widehat{LG}$. From the above 
identification we can see that the level is $-30$ which, if we assume 
the adjoint representation, is the negative of the dual Coxeter number 
of $E_8$, and so $G=E_8$. It is interesting that this identification
works for manifolds with a String structure. This is in
line with with the proposals in \cite{KS1, KS2, KS3} and
\cite{S1,S2,S3,S4} on the relevance of elliptic cohomology.

\vspace{3mm}
We distinguish between Dirac operators on the circle part and Dirac 
operators on the base part of the circle bundle. The study of the former 
uses Mickelsson's construction \cite{Mick}.
This then leads us to suspect the possibility of having a 
Wess-Zumino-Witten construction. Indeed we make the connection to such a 
construction, which suggests viewing spacetime as part of a generalized 
WZW model with $E_8$ as target. From the $LE_8$ point of view, the loop 
bundles coupled to the Dirac operator give contributions to the index. 
We also study the reduction of the $LE_8$ bundle down to finite 
dimensional bundles using \cite{CS} and interpret the corresponding 
Higgs field \'a la \cite{Hig}. 
An interesting example of the eta-form is the index gerbe \cite{Lo} 
related to the families index theorem \cite{CW}. We 
apply this to our problem and discuss the implications for string 
theory and M-theory. The latter is elaborated on in the last section.

\section{Review: Phase of the M-theory Partition Function} 
\label{MPF}
The topological part of the action that is used in the global 
M-theoretic considerations \cite{Flux, DMW, MS1, DFM} is the sum of the 
Chern-Simons term \cite{CJS} and the one loop term \cite{VW, DLM},  
\(
S_{11}=\frac{1}{6}\int_{Y^{11}} C_3 \wedge G_4 \wedge G_4
-\int_{Y^{11}}C_3 \wedge I_8,
\label{S11}
\)
where $I_8$ is the anomaly polynomial given in terms of the Pontrjagin 
classes of the tangent bundle of $Y^{11}$.

\vspace{3mm}
The above action was extended in \cite{Flux} to a 
twelve-dimensional manifold $Z^{12}$ whose boundary is the original 
eleven-manifold 
$Y^{11}$. This is possible because the relevant cobordism groups vanish.
In twelve dimensions, the action can be written in terms of the index 
of the Dirac operator coupled to $E_8$ and the Rarita-Schwinger 
operator, i.e. a Dirac operator coupled to $TY^{11}-3{\mathcal{O}}$
with $\mathcal{O}$ a trivial line bundle,
\(
S_{12}=\frac{1}{2}{\rm Index}(D_{V(a)}) + \frac{1}{4}{\rm 
Index}(D_{R.S.}),
\label{S12}
\) 
where $V(a)$ is the vector bundle associated to the $E_8$ principal 
bundle with characteristic class $a$ of dimension four.
Then, using the Atiyah-Patodi-Singer index theorem with the appropriate 
boundary conditions, the above action in eleven dimensions can  
be written in terms of the (reduced) eta-invariants, so that the 
resulting phase of the C-field is \cite{Flux}
\(
\Omega_M (C_3)= \exp\left[2\pi i \left(
\frac{{\overline\eta}(D_{V(a)})}{2}
+ \frac{{\overline\eta}(D_{R.S.})}{4}\right)\right].
\label{phase}
\)
The result is independent of the bounding manifold $Z^{12}$ used. 

\vspace{3mm}
We employ the geometric setup in \cite{MS1}. The Riemannian metric on
the circle bundle $Y^{11}$  is $g_{Y^{11}} =  \pi^{\ast}(g_{X^{10}})
+  \pi^*(e^{2 \phi/3}) {\cal{A}}\otimes  {\cal{A}} $,
where $g_{X^{10}}$ is the Riemannian metric on $X^{10}$,
$e^{2 \phi/3}$ is the norm of the Killing vector along $S^1$, which in
this trivialization is given by $\partial_{\theta}$, where $\theta$ is 
the coordinate on the circle, $\phi$ is the dilaton,
i.e. a real function on $X^{10}$ and  $ {\cal{A}}$ is a connection 
1-form on the circle bundle $Y^{11}$. Note that the component of the 
curvature in the direction of the circle action is
\(
R_{11}=e^{2\phi/3}=g_s^{2/3}.
\label{coupling}
\)
Such a choice of Riemannian metric is compatible with the principal
bundle structure in the sense that the given circle action acts as 
isometries on $Y^{11}$. Performing a rescaling to the above metric and 
using the identification
\eqref{coupling}, the desired metric ansatz leading to type  IIA is
\(
g_{Y^{11}}= g_s^{4/3} g_{S^1} + t g_s^{-2/3} g_{X^{10}}
\)
in the limit $t \rightarrow \infty$ then $g_s \rightarrow 0$.
\footnote{Note that we use the letter $g$ to denote both the metric 
(with a substript given by the symbol for a manifold) and the string 
coupling (with a subscript $s$). We hope that this will be clear from 
the context.}

\vspace{3mm}
Now let us consider the $E_8$-coupled Dirac operator $D$ on $Y^{11}$.
Using the formalism of Bismut-Cheeger \cite{BC} (and Dai \cite{Dai}) 
the adiabatic limit of the reduced eta invariants in the phase 
is \cite{MS1} 
\(
\lim_{t\to \infty}\overline\eta(D_{V(a)}^t)
=\int_{X^{10}} \hat{A}\left({\cal{R}}^{X^{10}}\right)
\wedge {\hat{\eta}}_{{V(a)}}  + \Sigma
\label{limit}
\)
where we write $\Sigma$ collectively for the terms that include eta 
invariants of the 
Dirac 
operator on $X^{10}$ coupled to the vector bundle ${\rm ker}D_{S^1}$
as well as for the dimensions of certain kernels. We do not record these terms  
as we will not need their explicit form in this note.

\section{Identification of the $G$-Bundle and the Eta-Forms}
\label{identif}
We would like to see whether anything explicit can be said about 
the expression (\ref{limit}). In particular, we would like to understand
whether a meaning can be given to the components of the eta-forms
$\widehat{\eta}$ of the $E_8$ bundle. The general strategy that we
follow is to try to identify as much as possible with terms that
exist in type IIA string theory. However, we would also like to see 
whether starting from a $G$-bundle we can {\it discover} that the 
structure group is $E_8$. This is what we would like to achieve.
Let us assume a general vector bundle associated to a principal 
$G$-bundle with an unspecified structure group given by a Lie group $G$.

\vspace{3mm}
Since the adiabatic limit involves the $\widehat{A}$-genus of the 
tangent bundle of the base $X^{10}$, from the point of view of type 
IIA this implies that we should seek terms that contain such 
gravitational terms. Indeed there is the one-loop term which has a 
degree eight gravitational piece. Thus we look at this term
\(
\int_{X^{10}} B_2 \wedge I_8,
\label{BI}
\)
which results from the reduction of the corresponding term in eleven 
dimensions and involves the B-field and the degree eight polynomial in 
the Pontrjagin classes of the tangent bundle of $X^{10}$
\(
I_8=\frac{1}{48}[p_2 - \lambda^2].
\)
It is obvious at this stage that (\ref{BI}) cannot be identified as it
stands with the piece with the same degree for the gravitational term as
in (\ref{limit}). However, this is possible provided that some 
assumptions hold. Note that while $I_8$ is not exactly $\widehat{A}_8$, 
the two are related via (this is derived and used in \cite{DFM})
\(
I_8=-30 \widehat{A}_8 + \frac{1}{8}\lambda^2.
\)
It is very interesting that if $X^{10}$ is a $String$ manifold, i.e.
with $\lambda=0$, then the comparison of (\ref{BI}) with (\ref{limit})
leads to a formula for the degree two component of the eta form
\(
\widehat{\eta}^{(2)}=-30 B_2 + d \alpha_1,
\label{comp}
\)
where $\alpha_1$ is some one-form. There are several points to be 
made at this stage. First, there is the extra factor $d\alpha_1$ that 
makes the identification of $\widehat{\eta}^{(2)}$ with the $B$-field 
only valid up to this exact term. Classically, we are interested in the
action and consequently in the eta-form. However,
quantum-mechanically,
what matters is the exponential of the action in the path integral,
which in the case of M-theory gives (fractional powers of)
\(
e^{2\pi i\eta},
\label{e}
\)
and in relating to type IIA we would be interested in the adiabatic
limit of the function (\ref{e}) rather than the adiabatic limit of
$\eta$ itself. As functions on the circle can be viewed as closed
one-forms, this means that at the quantum level we are interested in
the differential of the eta-forms $d {\widehat \eta}$. Via the
identification we made above, this then means that we should
be looking at the $H$-field rather than the $B$-field, which seems
to be consistent with consideration from twisted K-theory
in the general case. Considering then 
the differential of (\ref{comp}) removes the ambiguity coming from the 
exact term, and we thus have  
\(
d \widehat{\eta}^{(2)}=-30 H_3. 
\)

\vspace{3mm} 
Thus in our context the two-form component of $\eta$ is a connection on a
gerbe.
However, there is still a factor of 30 and a minus sign that need
to be explained. In general, the factor multiplying the gerbe is 
related to the representation of the group used. 
In particular, for the adjoint representation that number would be 
the dual Coxeter number $h^{\vee}$ corresponding to the spin gerbe. In 
our case, we can see that 30 is just the value of the
dual Coxeter number for $E_8$!
\footnote{Of course this is not unique. There are other 
choices: $A_{29}=\frak{su}(30)$, $C_{29}=\frak{sp}(58)$ and $D_{16}=\frak{so}(32)$.
Of the three the last seems the most relevant. In any case we 
leave this for future investigation.}
 Thus, from matching the topological 
terms in the action we are able to discover that the structure
group involved in the original bundle in eleven (and twelve) dimensions 
is $E_8$, provided we specify the representation to be the adjoint
representation-- which seems to be the most natural choice for a gauge 
theory. Alternatively, if we assume that we start with 
an $E_8$ gauge theory, then the above procedure specifies the 
representation for that $E_8$ principal bundle giving the associated 
vector bundle in the adjoint representation.
We will elaborate on this later in section \ref{WZW}. 

\section{The Four-Form as an Index}
\label{4as}

In this section we would like to see whether the four-form $G_4$ can 
have interesting expressions in certain situations. In particular, we 
would like to see whether the fact that the topological action 
(\ref{S11}) is written as an index (\ref{S12}) in twelve dimensions
is reflected in the topological part of the membrane action 
being also written as an index. We start by embedding the M2-brane 
in eleven-dimensional spacetime $Y^{11}$. 
The understanding of the topology of both
theories requires the extension to a `coboundary', namely the membrane
to a bounding 4-manifold $X^4$ and M-theory to a bounding 12-manifold
$Z^{12}$. Of course one cannot just pull back an index,
\footnote{We cannot just pull back the value of the index but it is 
possible to pull back the index as a bundle.} but we 
can assume that the vector bundles on $Z^{12}$ get pulled 
back to $X^4$. This means that the the gauge part of the Atiyah-Singer 
index theorem containg the Chern character will be the same. We then 
have to look at the effect of the gravitational term. In comparing  
$\widehat{A}(Z^{12})$ and $\widehat{A}(X^{4})$, it is obvious that 
they are in general different. However, they can give the same 
expression-- when integrated-- in a special case
on which we focus. Assume that $Z^{12}$ is decomposible into a product
of two spaces, a four-dimensional space which we identify 
\footnote{We think of the membrane as wrapping a subspace of 
spacetime.}
with the  
space $X^4$ cobounding the membrane, and an extra eight-dimensional 
piece $N^8$. The index in twelve dimensions would then decompose as
\(
\int_{Z^{12}} \widehat{A}(Z^{12})\wedge ch(E)=
\int_{X^4 \times N^8} \widehat{A}(X^4) \wedge 
\widehat{A}(N^8)\wedge ch(E),
\)
where we use the multiplicative property of the $\widehat{A}$-genus
$\widehat{A}(X_1 \times X_2)=\widehat{A}(X_1) \wedge \widehat{A}(X_2)$.
We get the desired result if we further assume that 
$N^8$ has $\widehat{A}(N^8)=1$. Such manifolds $N^8$ are called 
{\it Bott manifolds}, examples of which are manifolds of special 
holonomy. These manifolds occur naturally in compactifications of 
M-theory and string theory and so the situation that we described, 
although not completely general, is fairly generic in existing 
examples. Next we describe a different --but related-- way of getting 
an index expression for the form-field.

\subsection{The index gerbe}
\label{indexgerbesec}
In this section we consider a special kind of gerbe, namely the index
gerbe \cite{Lo, CW} .  Two motivations for this are the fact that the general 
formula for $d{\widehat{\eta}}$ is given by the integral over the fiber of an index,
and the embedding argument we gave in the preceding paragraph. 
A further motivation for considering this kind of 
gerbe is the following. We would like to understand
the effect of the Dirac operator of the vertical tangent bundle
on the eta invariant and consequently on the phase of the M-theory
partition function. Furthermore, we would like to understand the
local behavior, i.e. the behavior on local patches that cover the base,
and how these patch together to form the global objects that appear in
the eta invariant and in its adiabatic limit. If we concentrate on the
behavior of the phase of the Dirac operator of the vertical tangent
bundle on the local patches, then we are naturally led to the index
gerbe.

\vspace{3mm} 
The mathematical construction is given in \cite{Lo}, which we 
follow. For the M-theory circle bundle $Y^{11}$ with a 
projection $\pi$ to the type IIA base $X^{10}$, we consider the 
vertical tangent bundle
\footnote{Note that we use this notation not to mean the tangent bundle
of the circle itself.}
$TS^1={\rm ker}\hspace{0.5mm}\pi$ which can be viewed as a line
bundle over $X^{10}$. We assume that $TS^1$ has a spin structure and 
so we can form the corresponding spinor bundle $S(S^1)$. We also have 
$V$, a complex vector bundle on $Y^{11}$ with a compatible connection. 
For us, $V$ is either the vector bundle associated to the $E_8$ vector
bundle or the vector bundle $TY^{11}-3{\mathcal{O}}$, i.e. the
Rarita-Schwinger bundle. In this note we will concentrate on the first
of the two bundles, and so we will use $V$ to denote the $E_8$ bundle.
We couple the spinors on the vertical bundle to the vector bundle $V$
by forming $E=S(S^1)\otimes V$. Then $\pi_*E$ is the
infinite-dimensional vector bundle on $X^{10}$ whose fiber over $x\in
X^{10}$ is the space of sections $C^{\infty}\left( S_x^1;
E|_{S_x^1}\right)$. The base $X^{10}$ acts as a parametrizing
space for a family $D=\left\{ D_x\right\}_{x \in X^{10}}$ of Dirac
operators with $D_x$ acting fiber-wise on the space of sections over
$x$.

\vspace{3mm}
The construction of the index gerbe is as follows \cite{Lo}. We cover 
$X^{10}$ by
a set of charts $\{U_{\alpha}\}$ where $\alpha$ (and $\beta,
\cdots$) take values in an indexing set $I$.
The Dirac operator $D_{\alpha}$ defined over a patch $U_{\alpha}$
will have a modulus and a phase, the latter being given by
\footnote{Assuming no zero modes. An alternative description can be
found in \cite{CMM2}.}
\(
\frac{D_{\alpha}}{|D_{\alpha}|}.
\)
This is a mod 2 quantity, i.e. it takes the values $\pm 1$. We are
interested in the difference of phases on the overlap of patches. If
$U_{\alpha} \cap U_{\beta}$ is non-empty then the eigenvalue of the
operators
\(
\frac{D_{\alpha}}{|D_{\alpha}|} - \frac{D_{\beta}}{|D_{\beta}|}
\)
on the overlap can be $0,2$, or $-2$.

\vspace{3mm}
The components of the differential of the eta form are given as 
the corresponding components of the integral over the circle (i.e. 
the fiber) of the Atiyah-Singer index formula for the Dirac operator 
on $S(S^1)$ coupled to the vector bundle $V$, i.e.
for the coupled bundle $E$ \cite{Lo}. We are interested in the lowest
two degrees of the eta form, namely the degree zero and degree two
components.

\subsubsection{The zero-form component}
In this case the eta form is just half the Atiyah-Patodi-Singer eta
invariant. For the Dirac operator $D_{\alpha}$ over a patch
$U_{\alpha}$, this is given by
\(
{\widehat{\eta}}_{\alpha}^{(0)}=\frac{1}{2}\eta_{\alpha},
\)
so that the phase is simply
\(
\exp[2\pi i
{\widehat{\eta}}_{\alpha}^{(0)}].
\label{zero}
\)
On a nonempty interesection $U_{\alpha} \cap U_{\beta}$ the difference
\(
{\widehat{\eta}}_{\beta}^{(0)}|_{U_{\alpha} \cap U_{\beta}} -
{\widehat{\eta}}_{\alpha}^{(0)}|_{U_{\alpha} \cap U_{\beta}}
\label{rence}
\)
is a $\Z$-valued function. If $f_{\alpha}:U_{\alpha} \longrightarrow
S^1$ is defined by (\ref{zero}) then on the nonzero overlap
\(
f_{\alpha}|_{U_{\alpha} \cap U_{\beta}}=
f_{\beta}|_{U_{\alpha} \cap U_{\beta}},
\)
so that these functions $\{f_{\alpha}\}_{\alpha \in I}$ piece
together to form a function $f:X^{10} \longrightarrow S^1$, such that
the restriction of $f$ to a patch $U_{\alpha}$ is $f_{\alpha}$.

\vspace{3mm}
The differential of the eta form in this case is equal to the one-form
part of the Atiyah-Singer formula \cite{Lo}
\(
\frac{1}{2\pi i}d \ln f=\left( \int_{S^1}
{\widehat A}(R^{TS^1}) \wedge ch(F^V)
 \right)^{(1)} \in \Omega^1(X^{10}),
\)
which when evaluated gives
\(
\int_{S^1} c_1(F^V).
\)
In our case of an $E_8$ bundle, the first Chern class of the $E_8$ 
bundle is zero because
such bundles are characterized by a degree four class. This implies that
$f$ is constant, or more precisely that it is $\exp[2\pi i c]$ for a
constant $c$. 

\subsubsection{The two-form component}

In this case, what is important is differences of eta-forms on double
overlaps, and here one gets a gerbe \cite{Lo}. The line bundle that 
enters the
gerbe data is built out of the two eigenspaces with non-zero
eigenvalues, namely
\(
L_{\alpha \beta}=\Lambda^{\rm max} P_{+} \otimes \Lambda^{\rm max}
P_{-},
\)
where $P_{\pm}$ are the images of the orthogonal projections
onto eigenspaces with eigenvalue $\pm 2$. These images are
finite-dimensional vector bundles and so one can form the highest
exterior powers. On the images, $D_{\alpha}$
is positive and $D_{\beta}$ is negative on $P_{+}$, and
$D_{\alpha}$ is negative and $D_{\beta}$ is positive on $P_{-}$.
Furthermore, $L_{\alpha \beta}$ has a connection $\nabla_{\alpha
\beta}$ induced from the connections $P_{\pm}\nabla P_{\pm}$, and with
curvature $F_{\alpha \beta}$ an imaginary-valued two-form on
$U_{\alpha} \cap U_{\beta}$. In this case, the analog of 
(\ref{rence}) i.e. the difference of the two-forms on the overlap
is given by
\(
{\widehat{\eta}}_{\beta}^{(2)}|_{U_{\alpha} \cap U_{\beta}} -
{\widehat{\eta}}_{\alpha}^{(2)}|_{U_{\alpha} \cap U_{\beta}}
=\frac{-1}{2\pi i}F_{\alpha \beta}.
\)
The curvature of the gerbe with connection thus obtained is similarly
given by the degree three part of the integral over $S^1$ of the
Atiyah-Singer index formula
\(
d{\widehat{\eta}}^{(2)}=\left( \int_{S^1}
{\widehat A}(R^{TS^1}) \wedge ch(F^V)
 \right)^{(3)} \in \Omega^3(X^{10}).
\label{3}
\)
Evaluating this expression gives
\(
\int_{S^1}
-\frac{\rank(V)}{24}p_1 (R^{TS^1}) + c_2(F^V),
\)
which as a rational cohomology class lies in the image of
integral classes $H^3(X^{10};\Z)$ in the rational cohomology
$H^3(X^{10};\Q)$ as was proved in general in \cite{Lo}.


\section{The Loop Group Description}

It has been proposed in \cite{AE} that the $E_8$ bundle in M-theory
gives rise to an $LE_8$ bundle in type IIA on $X^{10}$.
This was studied further in \cite{MS1} where the the corresponding 
classes of the bundles were identified. Starting from  principal $E_8$ 
bundle over $Y^{11}$, the dimensional reduction of the M-theory to type 
IIA gives a $LE_8$ bundle $P\p$ in ten dimensions, characterized by the 
3-form $H_3=\int_{S^1}G_4$. Due to the homotopy type of the Lie 
group $E_8$, principal $E_8$ bundles over $Y^{11}$ are classified by
a class $a$ in $H^4(Y^{11}, \Z)$. Then the class on $LE_8$ is 
$u=\pi_{*} a \in H^3(X, \Z)$, which was identified in \cite{MS1} 
with the Dixmier-Douady class $DD(LE_8)$. 
\footnote{We continue to take $c_1=0$ for the circle bundle 
as remarked in footnote 1.}
Over each point in the base, 
the space of sections is identified with the loop group $LE_8$, because 
it can be viewed as maps from the M-theory circle to $E_8$, since the
bundle over the circle is trivial. Further, the obstruction to 
lifting the $LE_8$ bundle $P$ to an $\widehat{LE_8}$ bundle 
$\widehat{P}$, covering $P$, is the Dixmier-Douady class. That is, such 
a lift is possible only when $H_3=dB_2$ \cite{MS1}.

\subsection{The gerbe via loop space}
The isomorphism classes of gerbes on $X^{10}$ form an abelian group
$\mathbb{G}$ with the product structure given by the product of gerbes.
These isomorphism classes are classified by the characteristic class
$u$ for the gerbe, which is a map from $\mathbb{G}$ to third integral
cohomology. What
is the relation of the gerbes on $X^{10}$ to objects on the loop space?
In general there is a transgression map $T$ that takes 
$\mathbb{G}(X^{10})$
to the isomorphism classes of line bundles $Line(LX^{10})$ on the loop
space \cite{Bun}.

\vspace{3mm}
At the level of characteristic classes, there is a compatibility of
transgressions, i.e. the transgression of the characteristic class
$u(\mathbb{G})$ of a gerbe $\mathbb{G}$ is the first Chern class of
the transgression of the gerbe $T(\mathbb{G})$. The latter
is a line
bundle over the loop space so that $c_1(T(\mathbb{G}))\in
H^2(LX^{10};\Z)$ is the first Chern class of this line bundle
obtained by the transfer on cohomology
\(
T~:~H^3(X^{10};\Z) \longrightarrow H^2(LX^{10};\Z),
\)
and further, the transgression of the curvature of the gerbe, i.e. of 
the $H$-field, matches the curvature of the above line bundle over 
$LX^{10}$
obtained by transgressing the gerbe \cite{Bun}.
\footnote{ This is also done in the presence of a connection.} 
One can 
actually go one more step and relate the
gerbe to the holonomy of a connection over the double loop space of
$X^{10}$ by factoring the transgression above with the Bismut-Freed
relation between the determinant line bundle on a space and the
holonomy on the loop space \cite{Bun}.

\subsection{Twist vs. twisted, based vs. unbased}
The subgroup of based loops of $E_8$ is $\Omega E_8$, which is defined as 
the space of maps $f$ from the circle to $E_8$ that preserve the 
identity, i.e. such that $f(1)=1$.
The relation between the group of based loops $\Omega E_8$ and the
unbased ones $LE_8$ is $LE_8=\Omega E_8 \rtimes E_8$, which can be seen 
from the split short exact sequence
\(
\Omega E_8 \longrightarrow LE_8 \longrightarrow E_8.
\label{omeg}
\)
The multiplication $E_8 \times \Omega E_8 \rightarrow LE_8$ is a 
diffeomorphism, with the inverse given by $LE_8 \rightarrow E_8 \times 
\Omega E_8$ which takes $f$ to $(f(1),f(1)^{-1}f)$, and the two maps are 
smooth because of the differentiable structure on $\Omega E_8$.  
Using the inclusion $E_8 \hookrightarrow LE_8$, one can identify within 
the
class of $LE_8$-bundles those which come from $E_8$-bundles. For
example, within the class of associated vector bundles over $X^{10}$ with fiber
$\CC^{248}$ lie the vector bundles of the form $E \otimes L\CC$ with
$E\longrightarrow X^{10}$ a $248$-dimensional vector bundle.

\vspace{3mm}
There is a similar sequence of classifying spaces corresponding to 
(\ref{omeg})
\(
E_8 \longrightarrow EE_8\times_{\rm conj}E_8 \longrightarrow BE_8,
\)
with the indicated conjugation action.
Given a classifying map $X^{10} \longrightarrow BE_8$ one can then
pull back the $E_8$-bundle
\footnote{which is a bundle of groups and not a 
principal bundle.} 
over $X^{10}$. This can be interpreted as a bundle
over $X^{10}$ with fiber $B\Omega E_8$. Thus a section of this bundle
defines a {\it twisted} principal $\Omega E_8$-bundle over $X^{10}$.
A section of the $B\Omega E_8$-bundle is also a map from $X^{10}$ to
$EE_8\times_{\rm conj}E_8=BLE_8$ (with conjugation action) and thus
classifies principal
$LE_8$-bundles.

\vspace{3mm}
Conversely, given a classifying map $X^{10}\longrightarrow BLE_8
=EE_8\times_{\rm conj}E_8$, one can project down to $BE_8$ and
pull back the $E_8$-bundle as above. The original classifying map
defines a section of this $E_8$-bundle, and so a {\it twisted
principal} $\Omega E_8$-bundle over $X^{10}$. Hence {\it a principal
$LE_8$-bundle can be interpreted as a principal $\Omega E_8$-bundle
twisted by a principal $E_8$-bundle}. This is a new angle on the
construction in \cite{MS1} and is related to the nonabelian gerbe 
construction in \cite{AJ} for the case of the M5-brane. 

\vspace{3mm}
The usual viewpoint on the relation between the RR and the NSNS fields
is that the latter act as a twist to the former when described by 
cohomology or K-theory. The above bundle description, 
however, gives an alternative point of view where the NSNS fields 
seem to be the fields twisted by (part of) the RR fields. Furthermore, 
this provides some further justification--at least morally-- for 
the proposal in \cite{KS2} for treating the NSNS field
$H_3$ and the RR field $F_3$, in type IIB string theory, democratically,
that is, untwist the NSNS twist and view both fields as untwisted
elements of elliptic cohomology. In the current context, it is even more
because the twist is done by the RR field $F_4$, representing the
$E_8$ bundle, and what is being twisted is the NSNS field
$H_3$, representing the $LE_8$-bundle. The $E_8$-bundle that defines the
twisting is the pull-back of the {\it principal} $E_8$-bundle from
$BE_8$. The $B\Omega E_8$-bundle used above is the adjoint bundle of the
principal $E_8$-bundle.

\subsection{The String class}
In the case of the loop group, the DD-class is in fact just the String 
class \cite{Hig} which can be understood as an obstruction on the loop
space of our spacetime \cite{Kill, CP}. For physics purposes, it is 
desirable to 
work geometrically and, whenever possible, identify representatives of 
cohomology classes. Ref. \cite{Hig} provided an explicit 
differential 3-form representative of the de Rham image of the string 
class in real cohomology, which is defined using a connection and a 
Higgs field for the loop group. Using this, the string class of our 
$LE_8$ bundle on $X^{10}$ will be the integral over the circle 
of the Pontrjagin class of the corresponding $E_8$ bundle over 
$Y^{11}=S^1\times X^{10}$. 

\vspace{3mm}
For the $LE_8$-bundle $Q$, the string class can be explicitly 
characterized as follows \cite{Hig}. 
The Higgs field $\Phi$ is considered as the map from $Q$ to the space of smooth 
sections $C^{\infty}([0,2\pi], {\e}_8)$ satisfying the transformation 
property 
\(
\Phi(pg)={\rm ad}(g^{-1}) \Phi(p) + g^{-1}\partial_{\theta}~g,
\)
for $g \in LE_8$ and $\theta$ the coordinate on the circle. With $A$ a 
connection on $Q$ with curvature $F$, the 
string class of $Q$ is represented in de Rham cohomology by the 
three-form \cite{Hig}
\(
\frac{-1}{4\pi^2}\int_{S^1} \langle F, \nabla \Phi \rangle d\theta,
\label{dR}
\)
where $\nabla \Phi=\nabla_A \Phi - \partial_{\theta}A$. We notice that 
if the Higgs field is gauge-covariantly constant, i.e.
\(
\nabla_A \Phi=d \Phi + [A, \Phi]=0,
\)
then (\ref{dR}) becomes
\(
\frac{1}{4\pi^2}\int_{S^1} \langle F, \partial_{\theta}A \rangle 
d\theta.
\)

\subsection{The Higgs field and the reduction of the $LE_8$ to $E_8$}
In gauge theory, (spontaneous) symmetry breaking occurs if the structure
group $G$ of the principal bundle $P$ over $X$ is reducible to a closed 
subgroup $K$. This means that there is a principal subbundle of $P$ with
structure group $K$. The necessary and sufficient condition for such a 
reduction to occur is that the quotient bundle admits a global section. 
There is a one-to-one correspondence between these global sections 
$\Phi$ of the quotient bundle $P/K$ over $X$ and reduced subbundles
$P^{\Phi} \subset P$. These sections $\Phi$ are treated physically as 
the Higgs fields corresponding to the symmetry breaking. This effect is 
a quantum effect which occurs when the Lagrangian is invariant under 
the symmetry group but the vacuum is not.

\vspace{3mm}
In the Kaluza-Klein reduction of gravity on $S^1$, if one retains the 
non-zero Fourier modes then the resulting symmetry group on the base is 
a Kac-Moody extension of the Poincar\'e group \cite{DD}. Although this 
is a symmetry of the Lagrangian, it is not a symmetry of the vaccuum, which 
in the absence of a cosmological constant is Minkowski space, and the 
surviving symmetry group is just the Poincar\'e group. For instance, 
the 
dilaton-- the `size of the circle'-- acts as a Goldstone boson 
associated with the spontaneous breakdown of global scale invariance. 
In (super)gravity, the massive modes are spin-2 particles. Likewise, in 
the gauge theory we expect the massive modes to correspond to massive 
gauge bosons.

\vspace{3mm}
Guided by the above discussion, we expect then that the symmetry in the 
$LE_8$ gauge theory will be broken if the vacuum does not respect that 
symmetry. It then seems reasonable to assume that the 
resulting group will be the finite-dimensional part, i.e. the Lie group 
$E_8$, after truncating the Fourier modes coming from the loops.
\footnote{There are conditions for such a Fourier decomposition to occur.
We will discuss this in the next section.} 
Corresponding to the symmetry breaking
\(
LE_8 \supset E_8
\)
there is a bundle reduction from $Q$ to the subbundle $Q^{\Phi}$ with 
structure group $E_8$. The Higgs field $\Phi$ will then be a section 
of the quotient bundle $Q/E_8$, which is an $\Omega E_8$ bundle.
At the level of representations, the Higgs field will then take values 
in the corresponding $\Omega {\e}_8$ bundle.
\footnote{We are assuming a particular situation. The general case is 
discussed in \cite{Hig}.}

\vspace{3mm}
From the point of view of the gauge theory on the $S^1$ fiber the 
space of gauge orbits is the classifying space 
$B{\mathcal{G}}_0=\mathcal{A}/{\mathcal{G}}_0$, i.e. the quotient of 
the space of connections $\mathcal{A}$ by the based gauge
transformations ${\mathcal{G}}_0$.
The group of gauge transformations is just $\Omega E_8$ and so the 
space of equivalence classes is just $E_8$. So we see that from this 
point of view, modding out by $\Omega E_8$ corresponds to removing 
redundant degrees of freedom of the gauge theory on the $S^1$ part of 
spacetime.

\vspace{3mm}
The discussions above on the symmetry breaking are 
also in line with the expectation from couplings and considerations
of energy scales. Since the tension of a solitonic object is 
$~1/{\alpha'}^2$ whereas that of a RR object (i.e. a D-brane) is 
$~1/{\alpha'}$, then when the coupling is lowered the NSNS objects
are more massive. This can also be seen from the complementary 
picture using the field strengths in the (effective) action.

\section{Bundles from Loop Space}

\subsection{Breaking the loop bundle to $U(n)$}
\label{break}
In \cite{DMW}, the explicit comparison between M-theory and K-theory was done
by making use of the embedding  $(SU(5)\times SU(5))/\Z_5 \subset E_8$.
A natural question then is what happens when we start with the loop 
bundle of $E_8$. We can think of this in two ways. First, we can 
`loop both sides', i.e. get $LU(n)$ bundles from the $LE_8$ bundle 
and in order to get finite-dimensional vector bundles we can  
break $LU(n)$ to $U(n)$.
\footnote{The reason we are considering $U(n)$ instead of $SU(n)$ will 
be explained in section \ref{Relating}.}
 Second, we can start with $LE_8$ and break 
it to $E_8$, and then break $E_8$ to the unitary group {\'a} la DMW
($SU(5)$ is sufficient in ten dimensions due to stability \cite{DMW}). 
Note that the two ways are somewhat related because the classifying 
space for $LU(n)$ is the same as the loop of the classifying space of 
$U(n)$, i.e. $BLU(n)\cong LBU(n)$. We have seen an outline of how
the second scenario would work. In the rest of this section we consider 
the first scenario, and use the methods of \cite{CS} to get 
the decomposition
\(
LE_8 \supset LU(n) \supset U(n).
\)

\vspace{3mm}
Consider {\it rank-$n$ loop bundles} \cite{CS}, which are 
infinite-dimensional
bundles whose structure group is $LU(n)$ and whose fibers are isomorphic
to the loop space $L\CC^n$. The classifying space for such a bundle is
the loop space $LBU(n)$, and so the bundle is classified by a map
$f_E : X^{10} \longrightarrow LBU(n)$. Let
$ev : LBU(n) \longrightarrow BU(n)$ be the evaluation map that 
evaluates a loop at $1\in S^1$. The underlying $n$-dimensional vector 
bundle $U(E)\longrightarrow X^{10}$ is the bundle classified by the 
composition $ev \circ f_E : X^{10} \longrightarrow LBU(n)\longrightarrow 
BU(n)$.

\vspace{3mm}
We consider two classes of examples: 
\footnote{The two classes are related, as we will see shortly.}
\begin{enumerate}
\item First, looping the bundle $E \longrightarrow X^{10}$ leads to
$LE \longrightarrow LX^{10}$, which is classified by the map
$f_E : LX^{10} \longrightarrow LBU(n)\simeq BLU(n)$. If $E$ is the
tangent bundle $TX^{10}$, then $LE$ is the tangent bundle $TLX^{10}$
of $LX^{10}$. 
\item Second, tensoring the bundle $E$ fiberwise with $L\CC$.
This corresponds to the map of classifying maps $[X^{10}, BU(n)]
\longrightarrow [X^{10},LBU(n)]$ induced by the inclusion
$BU(n) \longrightarrow LBU(n)$ as the space of constant maps.
\end{enumerate}

\subsection{Fourier decomposition}

We will start by relating the second class of examples in 
the previous subsection to the 
situation in DMW \cite{DMW}.
There, spinors on $Y^{11}$ that transform as $e^{-ik\theta}$ under 
rotations of the circle were identified with the spinors on $X^{10}$ 
with values in $\cL^k$, where $\cL$ is the complex line bundle whose 
bundle of unit vectors is the M-theory $S^1$-bundle. The coupling to the
positive and negative chirality spin bundles $S^+$ and $S^{-}$,
coming from the decomposition of the spin bundle on $Y^{11}$ as 
$S=\pi^*(S^{+})\oplus \pi^*(S^{-})$, is $S^+ \otimes \cL^k$ and 
$S^- \otimes \cL^k$, respectively. Then, after including the coupling 
to the vector bundle $E$, one has the spinors coupled to the product
bundles $E\otimes \cL^k$. The loop description of this is given by the 
second example above, where we replace $E\otimes \cL^k$ with
$E\otimes L\C$. 

\vspace{3mm}
The loop bundles admit a (fiberwise) Fourier
expansion analogous to that of $\CC^n$-valued functions on the circle
$S^1$. For $L\CC^n$, this can be viewed as a map $\varphi$ from $L\CC^n$
to $\CC[[z,z^{-1}]]\otimes \CC^n$, the ring of formal power series in
$z$ and $z^{-1}$ tensored with $\CC^n$.
One can further restrict to the
`positive loops'
$L_+\CC^n=\varphi^{-1}\left( \CC[[z]]\otimes \CC^n \right)$. These are 
the
boundary values of the holomorphic maps from the two-disk to $\CC^n$,
$f : \mathbb{D}^2\longrightarrow \CC^n$.
In the current ten-dimensional setting, the disk is naturally
interpreted as the fiber over type IIA with the bounding theory $Z^{12}$
as the total space as in \cite{MS}.

\vspace{3mm}
The Fourier decomposition of $L\CC^n$ works as follows \cite{CS}.
The group of positive loops $L_+\CC^n \subset L\CC^n$ has the 
interesting property of being invariant under multiplication by $z$, 
i.e. $zL_+\CC^n$ is a subset of $L\CC^n$ with codimension $n$.
The inclusion gives rise to a filtration
\(
\cdots \subset z^{-k}L_+\CC^n \subset z^{-(k+1)}L_+\CC^n
\subset \cdots \subset L\CC^n,
\)
 where the union
$\bigcup_{k} z^{-k}L_+\CC^n$ is a dense subspace of $L\CC^n$.
This is the Fourier decomposition of $L\CC^n$. 

\vspace{3mm}
Analogously, a Fourier decomposition of rank $n$ loop bundle $E 
\longrightarrow X^{10}$ is a subbundle $E_+\subset E$ such that 
$E=E_+ \oplus E_-$ with
$E_+=E_-^{\perp}$ and $E_+$ is invariant under multiplication by
an element $z$ in formal Laurent polynomials $\CC[z,z^{-1}]$,
$zE_+ \subseteq E_+$ of codimension $n$. The bundle theoretic analog
of the Fourier decomposition of $L\CC^n$ is the filtration
\(
\cdots \subset z^{-k}E_+\subset z^{-(k+1)}E_+
\subset \cdots \subset E_+,
\)
whose union $\bigcup_{k} z^{-k}E_+$ is a fiberwise dense subbundle
of $E$ \cite{CS}.

\subsection{Relating $X^{10}$ to $LX^{10}$: conditions for Fourier 
decomposition}
\label{Relating}
The first of the two classes of examples in subsection \ref{break}
can be examined with the use of rank-$n$ loop bundles.
For $E\longrightarrow X^{10}$ an $n$-dimensional complex vector bundle
(in our main case of interest, namely $E_8$, we have $n=248$) classified 
by the map $f_E$ above, let $LE \longrightarrow LX^{10}$ be
the {\it induced} rank-$n$ loop bundle over the loop space $LX^{10}$.
The fiber of $LE$ over $\gamma \in LX^{10}$ is the space of sections
of the pull-back of $E$ over the circle,
$LE_{\gamma}=\Gamma_{S^1}\left(\gamma^*(E) \right)$.
For example, when $E=TX^{10}$, $LTX^{10}$ is the infinite-dimensional
tangent bundle of $LX^{10}$. The tangent space over $\gamma$ is the
space of vector fields living over $\gamma$.

\vspace{3mm}
Note that a Fourier decomposition is a much stronger condition than
a polarization since the latter allows for some finite-dimensional
ambiguity \cite{PS,CS}.
The homotopy type of a map of based loop spaces
$\Omega f_E : \Omega X^{10} \longrightarrow \Omega BU(n)
\simeq U(n)$ can be obtained by taking the holonomy map of
a connection on $E$. If we require $LE$ to have a Fourier 
decomposition, then the corresponding condition on $E$ is that
it must admit a homotopy flat connection \cite{CS}. Thus for our
$E_8$ bundle to admit such connections would mean that the 
bundle is essentially trivial.

\vspace{3mm}
A rank-$n$ loop bundle $\cE \longrightarrow X^{10}$ admits a Fourier
decomposition if and only if the structure group of $\cE$ can be reduced
to $U(n)$, viewed as the subgroup of constant loops in $LU(n)$
\cite{CS}. This can be rephrased in terms of disk bundles.
Let $f : X^{10} \longrightarrow LBU(n)$ classify a loop bundle
$\cE \longrightarrow X^{10}$. Then $\cE$ admits a Fourier decomposition
if and only if there is a lift of $f$ to the space of maps
${\rm Map}\left( {\mathbb{D}}^2, BU(n) \right)$ from the two-disk
${\mathbb{D}}^2$ to the classifying space $BU(n)$. Again, for us, this
${\mathbb{D}}^2$ is the fiber of type IIA in the bounding theory
$Z^{12}$ of M-theory, in the spirit of \cite{MS}. Thus this means a
twelve-dimensional extension.

\vspace{3mm}
Any loop bundle $\cE$ that has a Fourier decomposition has the following
description \cite{CS}. $\cE$ has a Fourier decomposition if and only if 
it is isomorphic to $L\CC \otimes E$, where $E$ is the underlying
$n$-dimensional bundle over $X^{10}$. Thus this brings us back to the
first class of examples and to the description of the the Fourier 
decomposition in terms of $L\CC \otimes E$ explaining the mode
expansions in \cite{DMW}.

\subsection{Producing Fourier-decomposable loop bundles via deformation}

There is a process that produces Fourier-decomposable bundles from 
loop space \cite{CS} which we now describe for completeness. Using the 
evaluation map $ev : LX^{10} \longrightarrow X^{10}$, one can pull back 
bundles from the spacetime
to loop space. The parallel transport operator induced by 
a connection on $E \longrightarrow X^{10}$ can be interpreted as an 
automorphism, i.e. a gauge transformation, of the pull-back bundle 
$ev^*(E) \longrightarrow LX^{10}$.
Loop bundles $LE$ can be deformed by such
gauge transformations. Let ${\cG} (ev^*(E))$ be the gauge group of 
bundle automorphisms of $ev^*(E)$. For $X^{10}$ smooth and simply 
connected, there is a natural rank-$n$ loop bundle
\(
L^{\cG}E \longrightarrow {\cG}(e^*(E)) \times LX^{10},
\)
satisfying interesting properties. For $t \in {\cG}(e^*(E))$ a gauge 
parameter, let $L^{t}X^{10}$ denote the restriction of $L^{\cG}E$ to
$\{t\} \times LX^{10}$; then

{\bf (1)} For the identity gauge element, ${\rm id} \in {\cG}(e^*(E))$,
$L^{\rm id}X^{10}=LE  \longrightarrow LX^{10}$;

{\bf (2)} For $t_{\nabla_E}$, the parallel transport operator of a
connection $\nabla_E$ on $E$, the bundle $L^{t_{\nabla_E}}E
\longrightarrow LX^{10}$ admits a natural isomorphism of loop
bundles,
\(
L^{t}E \iso C^{\infty}(S^1, \CC) \otimes ev^*E,
\)
and hence admits Fourier decompostion.
Thus, starting from bundles $E$ on spacetime $X^{10}$ we can build
Fourier decomposable bundles by going to loop space and performing 
a gauge transformation as above.

\vspace{3mm}
Can the loop bundle be reduced to groups other than $U(n)$?
It turns out that the only compact subgroups of $LU(n)$ are
conjugate to subgroups of $U(n)$ \cite{Stac} and so $U(n)$ is the
largest compact subgroup. This is appropriate and is in line with \cite{DMW}
as it leaves no
ambiguity in getting unitary bundles.

\vspace{3mm}
We can also ask whether one could have started with an $\Omega E_8$ 
bundle instead of an $LE_8$ bundle and performed the Fourier 
decomposition procedure on that bundle. In doing so one gets $\Omega U(n)$ 
in the intermediate step. However, there are no compact subgroups of
$\Omega U(n)$ \cite{Stac}, and so one cannot connect to 
finite-dimensional bundles
the same way.

\subsection{The eta invariant of the horizontal Dirac operator}
In section \ref{indexgerbesec} we related the eta invariant of the 
vertical 
tangent bundle to the index gerbe via the eta form that appeared in 
the adiabatic limit. Here we would like to briefly consider the part
that is related to the spin bundle of $X^{10}$, i.e. the horizontal
part. This gives the contributions to the action and the phase from 
the loop sector.

\vspace{3mm}
In DMW \cite{DMW} the Atiyah-Patodi-Singer $\eta$-invariant was 
decomposed according to Fourier modes $e^{-ik\theta}$ of the circle as
\(
\eta=\sum_{k\in \Z} \eta_k,
\label{et}
\)
where $\eta_k$ is the contribution from states that transform as 
$e^{-ik\theta}$ under rotation of the circle.
From our discussion above, it is clear that the natural generalization 
of (\ref{et}) is to consider coupling the Dirac operator to loop bundles
$LE$, both for the Rarita-Schwinger and the $E_8$ parts. 

\vspace{3mm}
We have seen how the vector bundles $E$ can be replaced by the loop
bundle $LE$ in order to account for the looping. We have also seen 
how such loop bundles can then be Fourier decomposed resulting 
eventually in the breakdown to $E \otimes L\C$. The first stage would
give the general contribution from the `loop sector'. This connects 
nicely to \cite{KS1,KS3}, where one of the ways of justifying the 
appearance of elliptic cohomology was to propose the source of this 
looping as being the Dirac operators coupled to the loop bundles. 
The second stage is obtained if one further wants to get the Fourier 
modes. Thus the two-stage picture looks like
\footnote{We could of course also loop the spin bundle itself. However, we 
preferred to keep the discussion in this section brief as we hope 
to revisit this elsewhere.}
\begin{equation} 
\begin{CD}
\eta_{{}_{S\otimes E}}  @>{\rm Looping}>> \eta_{{}_{S\otimes LE}}
@>{\rm Fourier}>>\eta_{{}_{S\otimes E\otimes L\C}} \\
\end{CD}\end{equation}

\section{The Generalized WZW Description}
\label{WZW}
In the discussion of the adiabatic limit, it was important to study the
Dirac operator on the circle bundle. Had the circle bundle been trivial
then we would not have had to analyze the eta invariants, since in that
case a symmetry argument would show that the eta invariant
contribution to the phase vanishes \cite{DMW}. 
What we are interested in is
the effect of the nontrivial M-theory circle, where no symmetry
arguments can be used to extract the contribution of the eta invariant
to the phase.
Thus, essential in our discussion is the Dirac operator on the circle
bundle, or more precisely, the Dirac operator on the `circle part'.
\footnote{Of course this is an oversimplification in terminology because 
the circle bundle is not a product.}

\vspace{3mm}
The physical nature of the $E_8$ gauge theory in eleven dimensions is 
not understood. It does not seem to be a Yang-Mills theory -- see the 
discussion in \cite{ES, DFM}. In particular we do not know the degrees 
of freedom of this theory.
\footnote{This is perhaps not surprising as we also do not know
the degrees of freedom of M-theory itself.}
Nevertheless, the structure of the topological parts of the action
seems to indicate that having an index of $E_8$ implies that we have
a curvature $F_2$ of the bundle, and so the
corresponding vector potentials must be present. \footnote{Of course 
alternatively, the way to describe this theory, if
it exists, could be very different from the standard methods of
differential geometry.} With this assumption we can look at the
Dirac operators coupled to these potentials. We can then form the
space $\mathcal{A}$ of ${\e}_8$-valued one-forms on $S^1$, where each
point $A \in \mathcal{A}$ defines a Dirac operator $D_A$ in the space
$\mathcal{H}$ of square-integrable spinors twisted by some
representation of $E_8$.

\vspace{3mm}
The principal loop group bundle gives rise via a representation to an 
associated vector bundle. The identification in section \ref{identif} 
indicates 
that we are dealing with the adjoint representation. Thus we represent 
$LE_8$ on its Lie algebra $L{\e}_8$ and consider the Hilbert space
$L^2(S^1,{\e}_8)$. We consider the complex Hilbert 
space $\cH$ that carries an irreducible unitary highest weight 
representation of the central extension $\widehat{LE}_8$ of the loop 
group $LE_8$ of level $k$. One has the Fock space as a product of a 
bosonic and a fermionic Fock space,
\(
{\cF}={\cF}_B^{(k,\lambda)} \otimes {\cF}_F^{(h^{\vee},\rho)},
\label{doublefock}
\)
labelled respectively by $\lambda$ and $k$, the weight of $E_8$ and 
the level, and by $h^{\vee}$, the dual Coxeter number and $\rho$, half the 
sum of
the positive roots. The dual Coxeter number is given by the value of the 
quadratic Casimir of $E_8$ in the adjoint representations, i.e.
\(
-2h^{\vee} \delta^{ab}=\lambda^{acd}\lambda^b{}_{cd},
\)
which has the value $30$ that we used in section \ref{identif}. 
Here the $\lambda$'s are the structure constants of $E_8$.
The Fock space is (\ref{doublefock}) with $k=-h^{\vee}$.

\subsection{The bosonic sector}
\label{bose}
The fact that the level in our case is given by the negative of the dual 
Coxeter number
implies that there is no bosonic Fock space and only ${\cF}_F$ occurs. 
The bosonic Fock space ${\cF}_B$ would correspond to the Sugawara currents.
The appearance 
of the dual Coxeter number $h^{\vee}$ implies that we are working at  
the {\it critical level} $k=-h^{\vee}$. The corresponding Kac-Moody 
symmetry is at the critical level and 
associated to it is a special `conformal field theory' which is not 
conventional because it does {\it not} have a stress tensor. It is in fact non-conformal.  

\vspace{3mm}
Let $\{J^a\}$ be the basis for ${\e}_8$, and 
$\{J_a\}$ the dual basis with respect to the Killing form, normalized so
that the length of the longest root is 2. The 
Sugawara-Segal current is 
\(
S(z)=\frac{1}{2}:J_a(z)J^a(Z):\sum_{n \in \Z} S_n z^{-n-2}.
\)
The commutation relations are
\(
[S_n, J_m^a]=-(k+h^{\vee}) m J_{n+m}^a,
\label{sj}
\)
\(
[S_n, S_m]=(k+h^{\vee})\left( (n-m)S_{n+m} + \frac{1}{12}k.{\rm dim} 
{\e}_8. 
\delta_{n,-m} \right).
\label{ss}
\)

\vspace{3mm}
Away from the critical level, i.e. when $k \neq -h^{\vee}$,
if the operators $S_n$ are scaled to $L_n=(k+h^{\vee})^{-1}S_n$ then
(\ref{ss}) generates the Virasoro algebra with central charge 
$c_k=\frac{k \cdot {\rm dim}\hspace{0.5mm}{\e}_8}{k+h^{\vee}}$. The replacement of $S_n$ by 
$L_n$ 
in (\ref{sj}) gives the action of the Virasoro algebra on 
$\widehat{L\e}_8$,
\(
[L_n, J_m^a]=-mJ_{n+m}^a.
\)

\vspace{3mm}
However, if $k=-h^{\vee}$ then the operators $S_n$ commute with 
the affine algebra and commute among themselves
\(
[ S_n, {J_m^a} ] = 0,
\label{SJ}
\)
\(
[ S_n, S_m ] = 0.
\label{SS}
\)
In particular, there is no usual conformal symmetry, and the second
relation 
implies the absence of the energy-momentum tensor.
If one was to mimic the construction for $k\neq -h^{\vee}$
then one would get that $L_n \rightarrow \infty$ and $c_{-h^{\vee}}
\rightarrow \infty$, as well as the commuting algebra.

\subsection{The fermionic sector }

We have just seen that the usual Sugawara currents are absent. Thus 
from here on we focus on the fermionic sector,
\(
{\cF}={\cF}_F^{(h^{\vee},\rho)}.
\label{FF}
\)
We follow \cite{Mick} where this construction was made (for a
different purpose). 
This (\ref{FF}) is the Fock space for the algebra of canonical 
anti-commutation 
relations (CAR) generated by the generators $\psi_n^a$, where $n\in \Z$ 
is a label for the momentum along the circle, and $a$ belongs to an indexing 
set $1,2,\cdots,\dim E_8=248$, that satisfy the canonical 
anti-commutation relations (CAR)
\(
\left\{ \psi_n^a,\psi_m^b \right\}=2\delta_{n,-m} \delta_{a,b}.
\)

\vspace{3mm}
The Fock vacuum is characterized by the zero mode Clifford subalgebra
of the CAR, and is a subspace of ${\cF}_F$ of dimension $2^{\dim 
E_8/2}=2^{124}$. This vacuum subspace carries an irreducible 
representation of the Clifford algebra generated by the $\psi_0^a$'s.
Any vector in the vacuum is annihilated by all $\psi_n^a$'s with $n<0$.

\vspace{3mm}
The generators $J_n^a$ of the loop algebra act on the Fock space 
${\cF}_F$ and they satisfy the commutation relations (CR's)
\(
\left[J_n^a, J_m^b \right]=-\lambda_{abc}J_{n+m}^c - \frac{h^{\vee}}{4}n 
\delta_{n,-m} \delta_{a,b}~,
\)
where $\lambda_{abc}$ are the structure constants on the Lie group 
$E_8$. Explicitly, the loop generators are given as bilinears in the 
oscillator generators
\(
J_n^a=-\frac{1}{4}\lambda_{abc} \psi_{n+m}^b \psi_{-m}^c,
\)
where normal ordering is not needed because the structure constants are 
totally antisymmetric. It is also understood that the RHS of this 
expression involves the sum over the contracted indices. 
The fermionic Hamiltonian is given by
\(
H_F=-\frac{1}{4}n:\psi_n^a \psi_{-n}^a: + 2h^{\vee}\cdot \frac{{\rm
dim}E_8}{24}
\)
with the reality condition $(\psi_n^a)^*= \psi_{-n}^a$.
The second term is the zero mode sector corresponding to the classical
case.
Corresponding to the Hamiltonian is its square-root, the supercharge 
$Q$, which satisfies $Q^2=H_F$, and which is defined by
\bea
Q&=&-\frac{i}{12} \lambda_{abc} \psi_n^a  \psi_m^b  \psi_{-m-n}^c
\nonumber\\
&=& \frac{i}{3}\psi_n^a J_{-n}^a. 
\eea

\vspace{3mm}
It is interesting to see what one gets when one restricts to the zero 
momentum mode sector. For this, the generators $\psi_n^a$ become the 
248-dimensional Euclidean gamma matrices $\psi_0^a=\gamma^a$ for the 
Lie group $E_8$. In this case, the supercharge $Q$ becomes (part of) 
Kostant's cubic Dirac operator 
\(
\cK=-\frac{i}{12}\lambda_{abc} \gamma^a \gamma^b \gamma^c.
\label{Kons}
\) 
Since the structure constants are totally antisymmetric, it follows
that the product of gamma matrices is totally antisymmetric, i.e. we
can replace the product of gamma matrices in (\ref{Kons}) by 
the antisymmetrized product $\gamma^{abc}$.

\vspace{3mm}
Since the dimension of the group manifold $E_8$ is even, we can define
the chirality operator $\psi_0^{249}$, the analog of $\gamma^5$ in four 
dimensions, and use it to get a grading operator
\(
\Gamma=(-1)^F \psi_0^{249},
\)
where $F$ is the fermion number operator and so $(-1)^F$ is the 
supersymmetry index. This gives, for $n\neq 0$,
\(
\psi_n^{a}F +F\psi_n^{a}=\frac{n}{|n|} \psi_n^{a}.
\)

\subsection{Coupling the supercharge to the vector potential}
\label{superch}
One can couple the supercharge operator to the vector potential
$A$ on the circle with values in the Lie algebra ${\e}_8$, to form 
the family of operators parametrized by $A$ \cite{Mick}
\(
Q_A=Q - \frac{h^{\vee}}{4} \psi_n^a A_{-n}^a.
\) 
where $A_{-n}^a$ are the Fourier components of $A$ satisfying
$(A_{n}^a)^*=-A_{-n}^a$. 
For a loop $g \in LE_8$, the corresponding lift $\hat{g}$ to the 
central extension $\widehat{LE}_8$ acts by conjugation on $Q_A$ 
resulting in a $g$-gauge transformation on $A$, 
\(
\hat{g}^{-1} Q_A \hat{g} =Q_{A^g},
\)
where $A^g=g^{-1}(A +  d)g$.

\vspace{3mm}
The operator $Q_A$ has a kernel that lies in the conjugacy class, so 
that $Q_A$ is not invertible on that set \cite{Mick}. This occurs at 
$\frac{\lambda+\rho}{k+h^{\vee}} \in {\h}^*$ in the dual Cartan 
subalgebra, which 
can also be viewed to be in the Cartan algebra $\h$ using the 
Cartan-Killing 
form. Our case is $k=-h^{\vee}$ and so this would imply that this set 
is infinite. However the situation is delicate.
\footnote{We thank Christoph Schweigert for an explanation on this.}
In general there is a continuum of conjugacy classes, which are given
for $SU(2)$ for example, by any latitude circle. The model however, 
picks out particular quantized conjugacy classes. In the classical 
limit a choice is $\lambda/k$, which can be seen as 
a limit of the general formula  $\frac{\lambda+\rho}{k+h^{\vee}}$ 
(cf. \cite{FFFS}). Since $k$ can be 
viewed as a measure of energy for the model then this latter 
equation is in some sense a strong coupling version of the classical 
formula. This can be seen for example by series expansion with 
$\lambda/k$ the lowest order 
term. One way to see that $k$ is a sort of momentum cutoff is by using 
the Peter-Weyl theorem, which says that $L^2(E_8)$ can be written as a sum 
over 
all representation labels $\gamma$ of the direct sum of the 
representation space $R_{\gamma}$ and its dual $R_{\gamma}^*$
\(
L^2(E_8)={\widehat{\bigoplus}}_{\gamma} R_{\gamma} \otimes R_{\gamma}^*.
\) 
In this context, the sum should be taken over $\gamma \leq k$ which 
gives the level $k$ the interpretation of a momentum cut-off. 

\vspace{3mm}
The family problem can be described by \cite{Mick}
\(
\begin{matrix}
{\rm Fred}_* &\longleftarrow & \mathcal{A} \cr
&&\dwn\pi' \cr
&& E_8 \cr
\end{matrix}
\)
where the top arrow takes a connection $A \in \mathcal{A}$ to the
corresponding supercharge $Q_A$ in the space of self-adjoint Fredholm
operators, and where the whole structure is over $E_8$. The vertical 
arrow relates $E_8$ to $\mathcal{A}$ via taking the holonomy of $A$, and 
the projection down to $E_8$, denoted by $\pi'$, is not always possible. 
The DD-class for $\pi$ is given by \cite{Mick}
\(
DD(\pi')=k \cdot \omega=-h^{\vee} \cdot \omega,
\)
where $\omega$ is the canonical integral generator of the cohomology 
$H^3(E_8;\Z)=\Z$ of $E_8$. 

\vspace{3mm}
At this stage the physics is occuring over the group manifold $E_8$ 
itself and not over the spacetime. Since $LE_8$ bundles over the 
ten-dimensional spacetime are completely 
characterized by the DD-class, then this means that we can pull back
from $E_8$ to spacetime and get our gerbes in spacetime to be 
coming from the basic gerbe on $E_8$. Since $LE_8$ bundles have a 
classifying space $E_8 \times BE_8$, then there is certainly a map
to $E_8$ viewed as part of the product. The discussion suggests 
a generalized WZW model,
i.e. with the two-dimensional worldsheet being replaced by the 
ten-dimensional spacetime.

\section{Further Discussion and Proposals}
We have identified the first two nonzero degrees of the eta forms 
by comparing the expression for the adiabatic limit of the eta 
invariant, representing the phase of the $C$-field in M-theory, with 
that of the topological term in type IIA string theory. The comparison 
gives essentially that $\widehat{\eta}^{(2)}$ is the $B$-field.
The construction works for String manifolds, i.e. for manifolds
with vanishing String class $\lambda=0$. The appearance of the dual 
Coxeter number in (\ref{comp}) can be used in two different ways. First,
starting from a $G$-bundle one discovers that $G$ should be $E_8$ 
under the natural assumption that the representation is the adjoint.
Alternatively, one can start with a specified $E_8$ gauge theory and 
using $h^{\vee}$ identify the relevant representation as being the 
adjoint representation. 

\vspace{3mm}
Further, we are able 
to use a generalized WZW construction to utilize the appearance of 
$h^{\vee}$, via the families index theorem and the twisted K-theory 
for $E_8$. Mapping to our families problem with Dirac operator on the
M-theory circle parametrized by points in the type IIA spacetime 
$X^{10}$, suggests the possibility of a generalized WZW model where the
spacetime is embedded in $E_8$. We also discuss the appearance of the 
infinite dimensional loop bundles, and their contribution to the 
partition function and the phase. To connect to finite-dimensional 
bundles the condition of Fourier decomposition arises, which puts 
severe restrictions on the bundle. We discuss the symmetry breaking 
problem in general for $LE_8 \longrightarrow E_8$ giving a Higgs field
with values in $\Omega E_8$. Corresponding to the infinite symmetry 
is an infinite number of generators, and when the symmetry is broken 
the generators are absent. This is another way to explain why in the 
construction only the fermionic Fock space was seen and the Sugawara 
bosonic currents were absent. 

\vspace{5mm}
{\bf 1. {\underline{The relation to twisted K-theory:}}}

\vspace{2mm}
Starting with a positive energy representation of $LE_8$ on
a Hilbert space $\cH$ (at a fixed level) one can take the principal
$LE_8$ bundle over $E_8$ and make it into a $PU(\cH)$ bundle using the
projective representation.
Using the construction in \cite{BCMMS}, twisted K-theory classes over
$E_8$ can be thought of as equivariant maps $f$ from $P$ to
${\rm Fred}_*$, where $P$ is the principal $PU(\mathcal{H})$ bundle over
$E_8$ with a given DD-invariant $\omega \in H^3(E_8;\Z)$, and
${\rm Fred}_*$ is the space of Fredholm operators, and the equivariance
condition is $f(pg)=g^{-1}f(p)g$ for $g \in PU(\mathcal{H})$.
In our case, the principal bundle $P$ is obtained by embedding the loop
group $LE_8$ inside $PU(\mathcal{H})$ through the projective
representation of $LE_8$
\cite{Mick}.
We have seen the relation to the twisted K-theory of the Lie group
$E_8$. However, we are ultimately interested in the twisted K-theory
of spacetime. Is there a way to get the twisted K-theory of spacetime
starting from the twisted K-theory of $E_8$? The mathematical answer
to this question is positive if there is a map from $X^{10}$ to
$E_8$ through which we can pull back the K-theory group. Physically, this
again suggests a generalized WZW sigma model $X^{10}
\hookrightarrow E_8$ ($=B\Omega E_8$). Alternatively, if there is a map to 
$E_8$ -- a section of the bundle-- one can use a pullback of the generator 
of the three-class in $E_8$ as the resulting $H$-flux in spacetime $X$,
and the pullback of $H^3(E_8)$ will give a subgroup of $H^3(X)$ isomorphic 
to $\Z$. This requires further investigation.
 
\vspace{5mm}
{\bf 2. {\underline{The sign involutions:}}}

\vspace{2mm}
Note that the identification of the two-form piece of the eta-form
included a minus sign in the prefactor. There are several sign reversals
that work nicely together. Reversing the sign of the Coxeter number
amounts to using the dual representation, i.e. lowest weight in place
of highest weight representation. For the gerbe itself, the reversal of
sign operation $B \mapsto -B$ has the following interpretation. At the
level of twisted Chern characters, this amounts to interchanging a
bundle $E$ with its complex conjugate $\overline{E}$, i.e.
\(
ch_{H}(E) \leftrightarrow ch_{-H}(\overline{E}).
\)
From eleven-dimensional supergravity, we have the gravitino supersymmetry rule
which gives the generalized spinor equation. Upon dimensional reduction
to ten dimensions the connection, built out of contractions of
$G_4$ with the eleven-dimensional gamma matrices, decomposes into two
connections, one for each of the positive and negative chirality spinor
bundles $S^{\pm}$, induced by the `generalized connection'
$\nabla \pm H$ of the tangent bundle, where $\nabla$ is the
Levi-Civita connection on the tangent bundle of $X^{10}$
and $H$ is the one-form built out of the NSNS field $H_3$. The choice
of sign corresponds to a choice of orientation, and so the reversal of
sign corresponds to the reveral of orientation of spacetime.
So the three ingredients: the dual Coxeter number, the twisted bundle,
and the twist work together. This means for instance that taking a
representation of highest weight using a twist $H$ for a bundle $E$ on
the spacetime $X^{10}$ with a given orientation is equivalent to
taking a lowest weight representation using a twist $-H$ for the
conjugate bundle $\overline{E}$ (coming from reversing the orientation).

\vspace{5mm}
{\bf 3. {\underline{The Ramond-Ramond fields:}}}

\vspace{2mm}
One might wonder whether the RR fields can be obtained from the
index gerbe and identified with the higher eta forms. While this would
be nice, it is unfortunately not the case. In principle,
from the index gerbe one can get even degrees on $X^{10}$ provided one
replaces the M-theory circle bundle with the bounding disk bundle, with
the corresponding expression given by replacing $S^1$ in the
Atiyah-Singer index formula giving the degrees $2k$ in the same
way that (\ref{3}) gives degree three. These are actually Deligne
classes \cite{Lo}. However, it is obvious from the index formula that it
contains the $\widehat{A}$-genus of the vertical tangent bundle and not
the (square-root) of the $\widehat{A}$-genus of $TX^{10}$. Had we been
able to get the RR fields this way (in cohomology) then we would have
obtained a derivation of the cosmological constant, i.e. the degree zero
component of the RR field, as the degree zero
component of the index formula. We now go back to the original situation 
with an odd-dimensional fiber. In the special case when the vertical
tangent bundle is trivial the degree zero component would be
\(
\int_{\mathbb{D}^2}c_1(V).
\)
For $E_8$, this vanishes, but can be non-zero for the
Rarita-Schwinger bundle.

\vspace{5mm}
{\bf 4. {\underline{Higher order corrections:}}}

\vspace{2mm}
The strategy we followed in the identification was to
use the topological terms that exist in type IIA string theory.
The rest of the terms that are not identified will either have
a different interpretation or that new terms in the type IIA action
have to be added to them to get a topological description for
the result.
In particular, we only focused on the $E_8$ part in M-theory and 
on the one-loop term in type IIA. The absence of the Chern-Simons term 
in our consideration of type IIA may be viewed as setting $F_4$ to zero. 
It would be interesting to pursue this
further. Higher terms should be explained in terms of higher order
corrections. However, at the moment these do not seem to be
written nicely in terms of characteristic classes.
\footnote{We thank B. Pioline for a comment on this. }
On the positive side, we expect the symmetry appearing to
turn out to be very useful in giving a handle on these higher
order (gravitational) corrections, e.g. by constraining their
structure.

\vspace{5mm}
{\bf 5. {\underline{The quantization conditions:}}}

\vspace{2mm}
Note that the index gerbe ended up being essentially the $H$-field.
Chasing this back to M-theory the standard
way we know that this comes from integrating $G_4$ over the circle.
This means that the index gerbe would come from $G_4$ written as an
index, and having the same expression (\ref{3}) without the integral.
Alternatively, if one uses the putative index formula in \cite{S1,S2}
then $G_4$ would have a shift coming from $\widehat{A}_4$ leading to
$G_4 - \lambda/24$. It is interesting that requiring this to be an
integral class implies that $\lambda$ is divisible by $24$, a condition
for orientability with respect to TMF. Note that this uses the
definition in \cite{S1,S2} for the zeroth component of the character to
be one, which in comparison can be seen to indicate the abelian nature
of $G_4$.

\vspace{5mm}
{\bf 6. {\underline{The role of $E_8$:}}}

\vspace{2mm}
We have started from a $G$-bundle in eleven dimensions and checked
whether one can {\it discover} $E_8$. Indeed if we specify the
representation to be the adjoint representation, which is the natural
choice for a gauge theory, then $E_8$ is specified via its dual Coxeter
number. Alternatively, if we choose to start from an $E_8$ bundle
in eleven dimensions then we can discover that the representation
for the loop group in ten dimensions is the adjoint. We believe that
this gives more evidence for the role of $E_8$ in M-theory, beyond
just being a model for $K(\Z,3)$ in low dimensions (compare 
\cite{Gauss,DFM}).
Another role of $E_8$ is suggested in section \ref{superch} where
the discussion indicates that spacetime may be used as the pre-image
of a generalized sigma model with $E_8$ as target.

\vspace{5mm}
{\bf 7. {\underline{The topology of spacetime:}}}

\vspace{2mm}
In section \ref{identif} we saw that the comparison of the adiabatic
limit of the eta invariant containing the eta-forms with the one-loop
term in type IIA string theory worked by imposing the String condition
(cf. \cite{Kill}, \cite{CP}) $\lambda=p_1/2=0$ on our manifold $X^{10}$.
This means that these manifolds are of special importance when
considering the global aspects of M-theory and string theory. This is in
line with with the proposals in \cite{KS1, KS2, KS3} and
\cite{S1,S2,S3,S4}. This should not be surprising since, after all, it 
is the topological one-loop term that led to the identification of the 
level as being the critical one.

\vspace{5mm}
{\bf 8. {\underline{Spacetime as parametrizing families:}}}

\vspace{2mm}
One interpretation of the viewpoint in the analysis of the Dirac 
operators in \cite{MS1} is that the ten-dimensional spacetime $X^{10}$ 
acted as a parametrizing family for the vertical Dirac operator, i.e.
the operator on the M-theory circle part. In view of the idea in this
paper this suggests a family version of the modularity problem.
This would encode the effects of the extra dimensions (over ten) in 
a systematic way.


\vspace{5mm}
{\bf 9. {\underline{Worldsheet vs. spacetime:}}}

\vspace{2mm}
In most of the comments in the other paragraphs the cited works on the 
critical level referred to that of the worldsheet conformal field 
theory. However what we have here is a CFT-like structure arising 
from spacetime, and more precisely from the extra directions over 
the ten-dimensional base. Is there any relation between the worldsheet 
on one side and the M-theory circle and the F-theory elliptic curve 
on the other? Indeed, it has been proposed in \cite{KS3} and further 
explained and elaborated in \cite{S4} that such a correlation exists.
Via this latter identification then what we are considering is a
`{\it CFT structure parametrized by type II spacetime}'.

\vspace{5mm}
{\bf 10. {\underline{ Adiabatic limits on the worldvolumes:}}}

\vspace{2mm}
The discussion in this note suggests that the structures on the 
spacetimes and on the worldvolumes are analogous and are correlated.
More precisely this implies that for the circle bundle $Y^{11}$ in 
the M-theory spacetime over the type IIA base $X^{10}$ there goes with 
it a corresponding circle bundle $M^3$ for the membrane over the string 
worldsheet base $\Sigma_g$. The worldsheet with the B-field in 
the action is the base of a $\mathbb{D}^2$-bundle giving the total
four-dimensional membrane cobounding theory with topological 
action the integral of $G_4$. Further, the discussion on the 
four-form being an index (section \ref{4as}) suggests that there is an 
adiabatic limit taken on the worldvolumes that corresponds to the one 
in the spacetime targets. Given the circle bundle $S^1 \to M^3 \to 
\Sigma_g$, with the metric 
\(
g_{M^3}=t g_{\Sigma_g} + A \otimes A,
\)
the the adiabatic limit for the vertical tangent bundle gives
\(
\lim_{t\to \infty}\overline\eta(D^t)=\int_{\Sigma_g} 
\widehat{A}(\Sigma_g) \wedge \widehat{\eta}.
\)
For dimensional reasons, $\widehat{A}(\Sigma_g)=1$ and so we are left 
with only the integral of the degree two eta-form 
$\widehat{\eta}^{(2)}$, which by our spacetime arguments is just 
$-30 B_2$. From the point of view of two dimensions this would then 
give some special role for theories in the large volume limit.

\vspace{5mm}
{\bf 11. {\underline{ The Higgs field and the topological membrane:}}}

\vspace{2mm}
Topological BF-theory for a G bundle on a Riemann surface $\Sigma_g$
(see e.g. \cite{Tho} for a review) is characterized by the flatness
condition $F_A=0$ as the extrema of the action. The moduli space for
such solutions is $\cM_F(\Sigma_g, G)$ and the corresponding field
theory has the partition function
\(
Z(\Sigma_g)= \int D \phi ~DA \exp \left(
\frac{1}{4 \pi^2} \int_{\Sigma_g} {\rm Tr}~ i \phi F_A \right),
\label{Fphi}
\)
where $\phi$ is the adjoint-valued field. In our case, we propose that
the corresponding theory will have the action (\ref{dR}) as the starting
point where the role of $\phi$ is played by $\Phi$ (in fact $\nabla
\Phi$ since we have a form degree shift). We interpret this as
giving the topological part of the membrane action where we have a
partition function analogous to (\ref{Fphi}).

\vspace{5mm}
{\bf 12. {\underline{The critical level:}}}

\vspace{2mm}
The supergravity description of string theory corresponds to the large
tension limit, i.e. when the string coupling $\alpha'$ tends to 0. A
consistent truncation is given by the massless modes which are the
fields of the effective theory.
We have seen in (\ref{bose}) that the theory we found is a peculiar
theory at the critical level, i.e. $k$ equals the negative of the dual
Coxeter number $h^{\vee}=30$. This theory is not conformal in the sense
that there is no energy-momentum tensor. The Segal-Sugawara current is
commutative. While there is no Virasoro symmetry one can still have the
affine Lie symmetry. Theories with
level $k$ equal to the dual Coxeter number are very special and they play
an important role in the geometric Langlands program since they provide
a natural way of constructing Hecke eigensheaves \cite{Lang}. The 
extensive construction of \cite{KW} does not make use of the critical level, 
nor of loop groups. \footnote{I thank Anton Kapustin for a comment on this.}
Thus our proposals can be seen as bringing in new elements in 
the connection between the physics and the mathematics,
albeit through a different approach.   
The relation (\ref{SJ}) implies that the operators $S_n$ belong to the
center of the enveloping algebra of $\widehat{L\e}_8$. This also
also implies that the algebra of infinitesimal diffeomorphisms of the
punctured disk acting on $\widehat{L\e}_8$ cannot be realized as an
internal symmetry of the space of states at $k=-h^{\vee}$.
What is playing the role of the curve in this case is the extra curve in
ten dimensions leading to F-theory as in \cite{KS3, S4}. This suggests
the current context as a potential setting for the Langlands program for 
ten-dimensional spacetime. We further note that from the
eleven-dimensional point of view, the infinite volume limit for
the ten-dimensional base-- which is desirable for the sigma-model 
(cf. \cite{FL})-- is just the adiabatic limit of the circle bundle, the main setting 
for this note.

\vspace{5mm}
{\bf 13. {\underline{High energy and tensionless limits of strings:}}}

\vspace{2mm}
What is the physical implication of the critical level? The answer is
that it provides a window for high energy regimes. In \cite{LZ} it was
argued that the critical level corresponds to the tensionless limit of
string theory. This is the limit where a huge new symmetry emerges due
to the dramatic increase in the number of zero-norm states. Classically,
the tensionless limit arises when $k \to \infty$ but quantum
mechanically one has to include the shift by $h^{\vee}$. One thus
expects that the shifted level to be a measure of coupling.
Indeed, the WZW analysis leads to the idenitification of the level as
an inverse coupling constant, namely
\footnote{The analysis in \cite{BS} is done for noncompact cosets. We
remove a relative minus sign (cf. entry {\bf 2} above). For
$k=-h^{\vee}<0$, the vacuum is given by
\bea
J_n^a|~0 ~\rangle&=&0 ~~~{\rm for}~~~n>0
\label{one}
\\
\psi_n^a |~0~ \rangle&=&0 ~~~{\rm for}~~~n>0.
\label{two}
\eea
If (\ref{one}) is used then (\ref{two}) is satisfied (also works for
both having $n<0$). Unitarity is fixed if the representation is flipped 
from positive energy to negative energy because we get the desired sign
in the commutators.}
\cite{LZ,BS}
\(
\alpha'=\frac{1}{k + h^{\vee}}.
\)
Thus the
critical level seems to be the appropriate setting for studying very
high energy properties of string theory where the huge symmetry is
still unbroken. The discussion in this note suggests likewise that
large spacetime symmetries can be detected and should be analyzed at
this level.

\vspace{5mm}
{\bf 14. {\underline{The non-commutative geometry of spacetime:}}}

\vspace{2mm}
For the WZW model, the classical level, i.e. when $k \to \infty$,
corresponds to classical geometry. As the level is brought back
from infinity to smaller and smaller values, the classical geometry
of spacetime starts undergoing quantum deformations that introduce
noncommutativity to the coordinates so that the spacetime becomes
noncommutative (see \cite{FG}). The commutativity keeps increasing
until the level reaches the critical value at which stage the
the geometry becomes singular and the
noncommutativity becomes infinite \cite{BS}. In our case of course
the B-field was essential in deriving the critical level. The
point is that in this limit the discussion of noncommutativity should
be nonperturbative as opposed to the usual perturbative
deformation approach that is done at finite values of the
noncommutativity, i.e. spacetime is intrinsically noncommutative
at the quantum level as opposed to being deformed to be so.

\vspace{5mm}
{\bf 15.{\underline{ The cosmological constant:}}}

\vspace{2mm}
Recall again that, in the WZW model, the classical limit corresponds to taking 
the level to infinity, $k \to \infty$. From a geometric point of view, taking 
the classical limit means going to the large volume (or `radius') limit, 
because the more the curvature increases the more quantum effects we 
have and the deeper we go into the quantum regime. One way of measuring 
this is through the cosmological constant $\Lambda_g$, whose value can 
also be a measure of curvature. Large values of $\Lambda_g$ correspond 
to high curvature and hence to strong coupling. From this we see that in 
type IIA string theory the cosmological constant, which is the zeroth 
component $F_0$ of the Ramond-Ramond field $F$, if related to the central 
extension of $LE_8$, as suggested in \cite{AE}, then such a relationship 
should be more of an inverse relationship rather than the linear 
relation $F_0=k$ that was proposed in \cite{AE}. This is because
for the same space -- the ten-dimensional type IIA manifold-- having $k 
\to \infty$ would then mean $F_0 \to \infty$ which gives strong 
coupling, in contrast to AdS/CFT where one is comparing the coupling of 
the bulk to the coupling on the boundary related by strong/weak duality. 
However, if {\it some} 
form of a one-to-one relationship 
exists between 
the cosmological constant and the central extension, then our discussions 
in this note would specify the cosmological constant because the 
theory singled out a particular level, namely the critical level.
It seems reasonable (e.g. from coset model considerations \cite{LZ}) 
to expect $F_0 =T/k$, where $T$ is the string tension.

\vspace{5mm}
{\bf 16. {\underline{Holography:}}}

\vspace{2mm}
The appearance of the $E_8$ Kac-Moody symmetries suggests the 
possibility that type IIA string theory have a description in  
terms of a quantum field theory. One can also go further to 
ask whether there is a holography in which type IIA is the
codimension one theory. Indeed if this is the case then we 
can suggest the cobounding theory of type IIA, i.e. the 
eleven-dimensional `theory' -- let us name it $M_A$-- on $M^{11}$ whose 
boundary is type IIA. This is the result of viewing the two-disk bundle 
$\mathbb{D}^2$ over type IIA in two different ways. Starting from 
the twelve-dimensional theory on $Z^{12}$ we can take its boundary
$Y^{11}$ to arrive at M-theory and then take the $S^1$-reduction 
to arrive at type IIA, or alternatively take the $S^1$ reduction 
of $Z^{12}$ to get the `theory' $M_A$ and then take the boundary to get 
to type IIA string theory. The topological terms would then be 
\(
\frac{1}{6} \int_{M^{11}} H_3 \wedge F_4 \wedge F_4 - I_8 \wedge F_4.
\) 
Further, this can also be seen as a more intrinsic definition 
of the differentials of the eta-forms $d\widehat{\eta}$. 

\vspace{3mm}
Obviously there is a lot of work to be done. We hope to gain 
a better understanding and to report more in the near future.

\bigskip\bigskip
\noindent
{\bf \large Acknowledgements}\\
\noindent
I would like to thank Alan Carey, Jarah Evslin, Varghese Mathai,  
Jouko Mickelsson, Tony Pantev, Christoph Schweigert, and  
Bai-Ling Wang for useful discussions and explanation, and Arthur Greenspoon 
for suggestions on improving the presentation.
I especially thank Jouko Mickelsson for very useful 
comments and suggestions on the manuscript, especially on section 
\ref{WZW}. I also thank Ralph Cohen for explaining 
the main result of \cite{CS}.
I acknowledge the hospitality of IHES, DIMACS and the Department of
Mathematics at Rutgers, MSRI and the organizers of the program 
``New Topological Structures in Physics" where  part of 
this work was carried out. This research is supported by an ESI
Junior Research Fellowship associated with the program "Gerbes, Groupoids, 
and Quantum Field Theory".

\noindent


\end{document}